\documentclass[final,twocolumn,nofootinbib,superscriptaddress,nobalancelastpage]{revtex4}
\pdfoutput=1

\usepackage{graphicx}
\usepackage{rotating}
\usepackage{amsmath,amstext,amsfonts,amsbsy,amssymb,amscd,bbm,epsfig,lscape}

\usepackage{hyperref}

\newcommand{\be}{\begin{equation}}
\newcommand{\ee}{\end{equation}}
\newcommand{\bea}{\begin{eqnarray}}
\newcommand{\eea}{\end{eqnarray}}

\newcommand{\Cr}{$^{51}$Cr\ }
\newcommand{\Rn}{$^{222}$Rn\ }
\newcommand{\Kr}{$^{85}$Kr\ }
\newcommand{\Xe}{$^{136}$Xe\ }

\newcommand{\twobeta}{$2\nu\beta\beta$~}

\begin{document}

\title{Combining dark matter detectors and electron-capture sources to hunt
  for new physics in the neutrino sector}

\author{Pilar Coloma}
\email{pcoloma@vt.edu}
\author{Patrick Huber}
\email{pahuber@vt.edu}
\author{Jonathan M.~Link}
\email{jmlink@vt.edu}
\affiliation{Center for Neutrino Physics, Physics Department, Virginia Tech, \\ 850 West Campus Dr, Blacksburg, VA 24061, USA}

\begin{abstract}
In this letter we point out the possibility to study new physics in
the neutrino sector using dark matter detectors based on liquid
xenon. These are characterized by very good spatial resolution and
extremely low thresholds for electron recoil energies. When combined
with a radioactive $\nu_e$ source, both features in combination allow
for a very competitive sensitivity to neutrino magnetic moments and
sterile neutrino oscillations. We find that, for realistic values of
detector size and source strength, the bound on the neutrino magnetic
moment can be improved by an order of magnitude with respect to the
present value. Regarding sterile neutrino searches, we find that most
of the gallium anomaly could be explored at the 95\% confidence level 
just using shape information.
\end{abstract}

\maketitle

\section{Introduction}

Neutrinos have long been a rich hunting ground for physics beyond the
Standard Model (BSM).  In fact, neutrino mass is so far the only
BSM physics that has been established in laboratory experiments.
Astrophysical evidence of dark matter suggests the
existence of BSM particles, which have nevertheless not been observed yet. Among all feasible candidates, weakly
interacting massive particles (WIMPs) are theoretically rather appealing. These may be
observable through their interactions within detectors, as the earth moves
through the sea of WIMPs. This possibility has triggered a cornucopia
of experimental efforts of direct dark matter
detection~\cite{Bauer:2013ihz}.

In this letter we examine the physics potential of combining a liquid
xenon (LXe) detector, designed to search for WIMP dark matter, with an
intense electron-capture neutrinos source in order to look for
neutrino magnetic moments ($\nu$MM) and other new physics in $\nu_e e^-$
elastic scattering. The idea of looking for new physics in the neutrino sector using dark matter detectors has been proposed before in the literature, see for instance Refs.~\cite{Pospelov:2011ha,Harnik:2012ni,Pospelov:2012gm,Pospelov:2013rha}. Direct dark matter detection relies on observing
nuclear recoils with electron-equivalent energy down to $\sim\!1$~keV. 
Due to the small values expected for the dark matter interaction cross 
section, large detector masses and low background levels are also required.
A LXe time projection chamber (TPC) can provide a large volume, low
detection thresholds (sub-keV) and a very low background rate at the
energies of interest.  At the same time the electron density is
higher in xenon than in any other stable noble gas, thus providing the largest
possible target density in any given volume near the source.  The idea
of using liquid noble gas detectors to search for $\nu$MM was
first suggested by Vogel and Engel~\cite{Vogel:1989iv}, but never
developed. As a by-product we also find non-negligible sensitivity to
sterile neutrino oscillations in the $\Delta m^2\sim$1~eV$^2$ range
suggested by recent terrestrial experiments~\cite{Abazajian:2012ys}.

When a nucleus decays via electron-capture almost all of the available 
energy goes into a mono-energetic neutrino. Among possible nuclei which decay 
via electron-capture, \Cr offers several practical advantages: it is readily 
produced by thermal neutron capture~\cite{Cribier:1996cq}, has a mean lifetime of
39.96~days and produces two mono-energetic neutrino lines at 750~keV
(90\%) and 430~keV (10\%). Mega-curie-scale \Cr sources have been
produced in the past and used to calibrate the gallium radiochemical
solar neutrino detectors GALLEX~\cite{Anselmann:1994ar,Hampel:1997fc}
and SAGE~\cite{Abdurashitov:1998ne}.  

\section{Constraints on the neutrino magnetic moment}

In the presence of a $\nu$MM, the differential cross section for
$\nu_e e^-$ elastic scattering can be written as
\begin{equation}
\left(\frac{d\sigma}{dT}\right)_{tot}\! = \!\left(\frac{d\sigma}{dT}\right)_{SM}\! + 
\frac{\pi \alpha^2}{m_e^2}\!\left(\frac{1}{T}\! -\!
\frac{1}{E_\nu}\right)\!\!\left(\frac{\mu_\nu}{\mu_B}\right)^2  ,
\label{eq:xsecmu}  
\end{equation}
where $m_e$ is the electron mass, $T$ is the electron recoil energy,
$E_\nu$ is the neutrino energy and $\mu_\nu$ is the $\nu$MM in Bohr
magnetons ($\mu_B$). The term proportional to $\mu_\nu$ produces an
increase in the number of events at low electron recoil energies.
This makes two-phase LXe
TPCs~\cite{Lebedenko:2009xe,Aprile:2011dd,Akerib:2012ys,
  Aprile:2012zx,Baudis:2012bc,Malling:2011va}, with their low-energy
detection threshold, ideal detectors for such a search. Currently, the
lowest bounds on $\nu$MM come from astrophysical
observations~\cite{Raffelt:1999gv}: $\mu_\nu \lesssim 3\times
10^{-12}\mu_B$. The best constraint from terrestrial
experiments, on the other hand, has been obtained by the GEMMA experiment, $\mu_\nu <
2.9\times 10^{-11}\mu_B$ at 90\% CL~\cite{Beda:2012zz}. In the SM,
$\nu$MMs are expected to be many orders of magnitude below present
bounds, yet many extensions of the SM produce an enhancement of
the $\nu$MM, see for instance Ref.~\cite{Broggini:2012df} and references therein.

\begin{figure*}[t]
  \centerline{\includegraphics[width=\columnwidth]{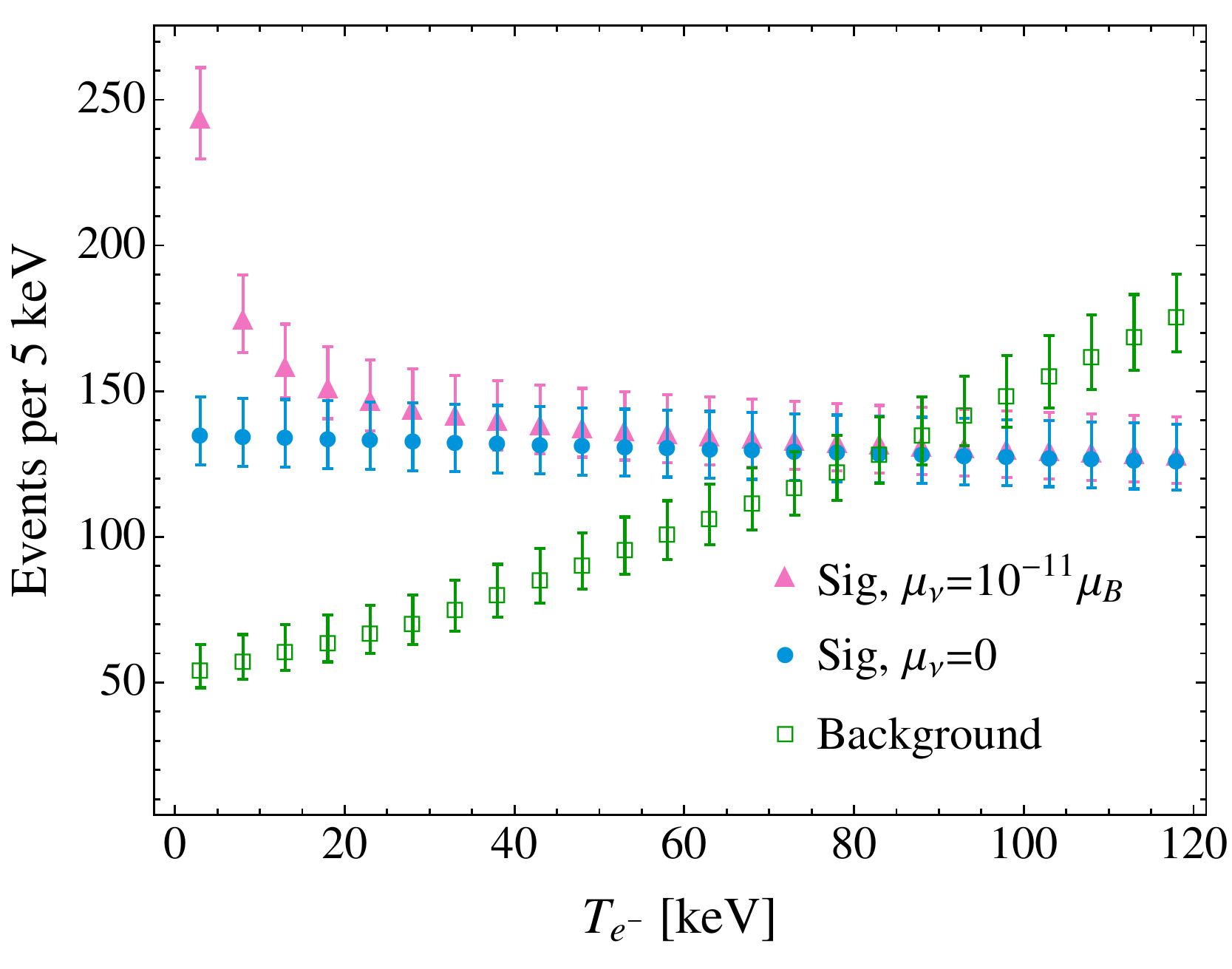}\hspace{2ex}%
  \includegraphics[width=\columnwidth]{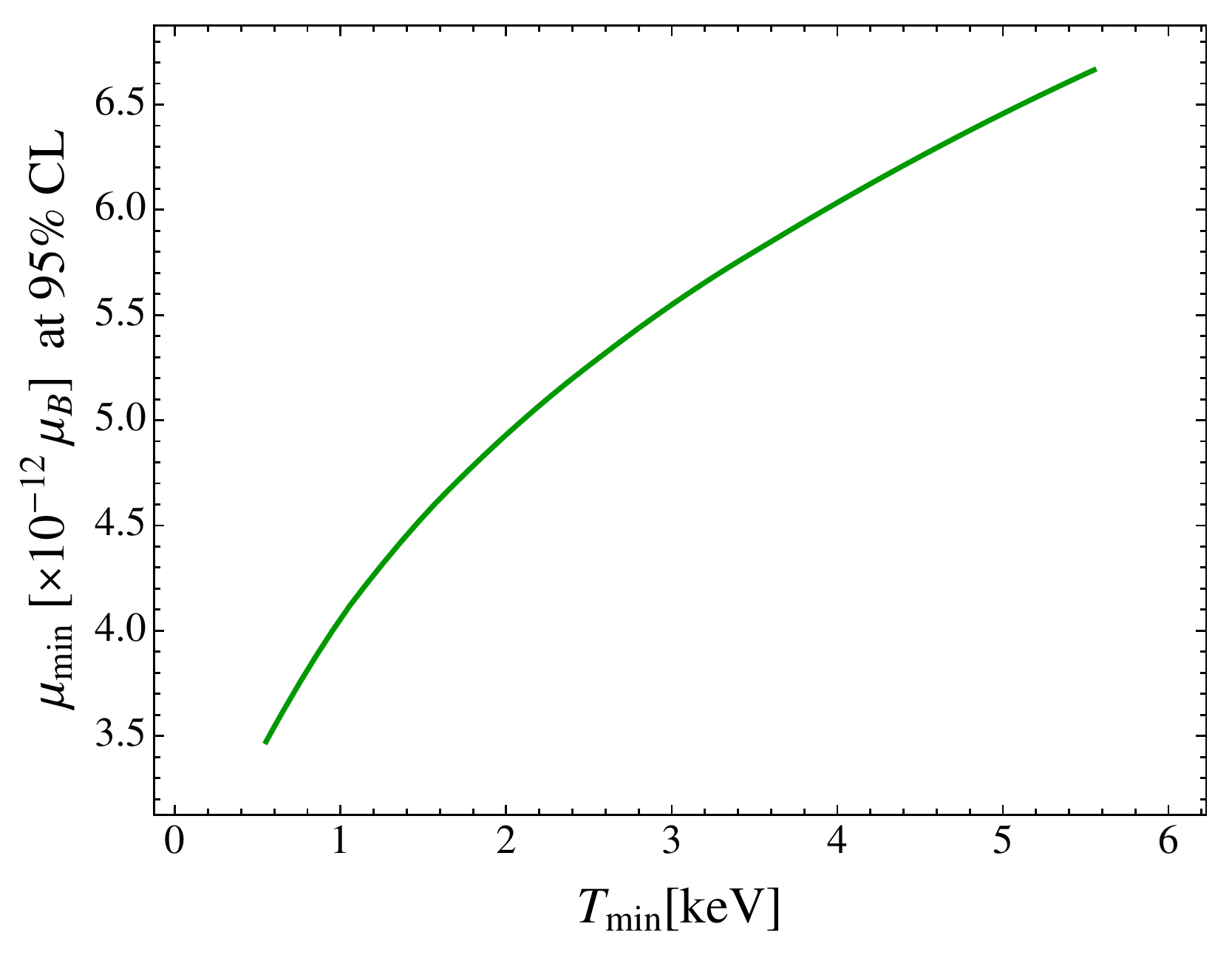}}
  \caption{Left panel: Number of events (per 5 keV) as a function of the recoil 
    energy of the electron. Results for the signal are shown with 
    (pink triangles) and without (blue circles) a $\nu$MM. The expected 
    background event rates are also shown for comparison (green squares). 
    Right panel: The achievable bound on the $\nu$MM (at the 95\% CL) as a function 
    of the low-energy threshold on electron recoil energy.}
\label{fig:mu}
\end{figure*}

For our sensitivity estimate, we assume a data taking period of 100~days, using 
a \Cr source with initial strength of 5~MCi. Our choice for the strength of the source is based on 
simulations conducted for the SOX experiment~\cite{Borexino:2013xxa} of the 
GALLEX enriched $^{50}$Cr material~\cite{Cribier:1996cq} irradiated in the High Flux Isotope
Reactor at Oak Ridge National Laboratory. We consider a generic LXe detector, 
but for definiteness we chose a design similar to the proposed LZ
detector~\cite{Malling:2011va,private}. We assume a cylindrical fiducial
volume with equal diameter and height, $h=1.38$~m, which contains
$\sim\! 6$~tons of LXe, and we assume that the radioactive source is
placed 1~m below the fiducial volume, along the central axis of the
cylinder. Neutrinos are detected via electron elastic scattering in
the detector, see Eq.~\ref{eq:xsecmu}. Under these assumptions, a 
total of $12,518$ signal events are expected for a 100~day
run. Regarding backgrounds, we have considered contributions from
solar neutrino interactions, \Rn and \Kr decay, and the \Xe~\twobeta
decay.  Following Ref.~\cite{Baudis:2013qla}, the solar neutrino
background is estimated to be 1.05 counts per ton and day for $pp$
neutrinos and 0.51 counts per ton and day for $^7$Be neutrinos. The
\Kr and \Xe backgrounds have been taken directly from Fig.~2 in
Ref.~\cite{Baudis:2013qla} and rescaled according to our run length
and detector mass, while for \Rn we assume that a goal of
0.1~$\mu$Bq/kg can be achieved~\cite{Baudis:2013qla}, about a
factor of 30 reduction with respect to what has been achieved for
EXO-200~\cite{Albert:2013gpz}. 

Finally, an important source of background could come from the source 
itself.  In 10\% of \Cr decays there is a 320~keV gamma, which can easily be shielded 
with just a few cm of tungsten. However, impurities present in the chromium prior to 
irradiation, can lead to the production of MeV gamma emitters~\cite{Bellotti:19965, 
Abdurashitov:1998ne}.  These will require significant additional shielding to be reduced down to
an acceptable level.  We base our calculation of this background on the measured gamma 
activity of the GALLEX source~\cite{Bellotti:19965}. Our source is assumed to be 
shielded by a 17~cm thick tungsten layer, which, when combined with 70~cm of LXe (present 
between the tungsten shield and the edge of the detector), provides an attenuation of 
$10^{-11}$ for a 1.5~MeV gamma. Nevertheless, with $10^{13}$ gammas emitted per day (in all 
directions), we expect about 10 to pass through the shielding and Compton scatter in the 
detector fiducial volume. Most of these would deposit energy in excess of the maximum 
from a \Cr neutrino, though. Further suppression could come from a veto on mulit-site Compton 
scattering events.  We estimate the surviving background in the fiducial volume from source 
gammas to be less than 1 event per day at the start of the data taking, and, since $\sim$95\% 
of MeV gammas come from short-lived isotopes (such as $^{64}$Cu, $^{77}$Ge and $^{24}$Na), 
this rate should rapidly decay with time. Therefore, we will neglect these events in our 
analysis, but we note that care must be taken in the preparation of the $^{50}$Cr source 
material to ensure the required level of purity is reached.

To constrain the $\nu$MM, the analysis is done in terms of the recoil
energy of the electron only. The recoil energies are smeared on an
event-by-event basis according to a Gaussian with $\sigma(T) =
0.20\sqrt{T}$. Generally, we find that the energy resolution does not
have a significant impact on the $\nu$MM sensitivity.  A Poissonian
$\chi^2$ is constructed using 0.1~keV wide bins in $T$, from 2~keV to
140~keV unless otherwise stated. In this energy range a total of $3,656$
signal events are expected, together with a total of $3,450$ background
events. The distribution of signal and background events in electron
recoil energy is shown in Fig.~\ref{fig:mu} (left). Reducing the backgrounds
does not significantly improve the sensitivity to this observable
since in the low-energy region, where the $\nu$MM enhancement is
expected, the background rate is already quite low. On the other hand,
we find the low-energy threshold to be the most relevant parameter in
this analysis (as expected from Eq.~\ref{eq:xsecmu}). Fig.~\ref{fig:mu} (right) shows the dependence of the
sensitivity as a function of detector threshold. With a fairly
conservative threshold of 2~keV the 95\% CL bound is $\mu_\nu < 4.9
\times 10^{-12}\mu_B$. Such values would yield an improvement of a factor of 5
over the currently best terrestrial limit.

\begin{figure*}[t!]
  \centerline{\includegraphics[width=\columnwidth]{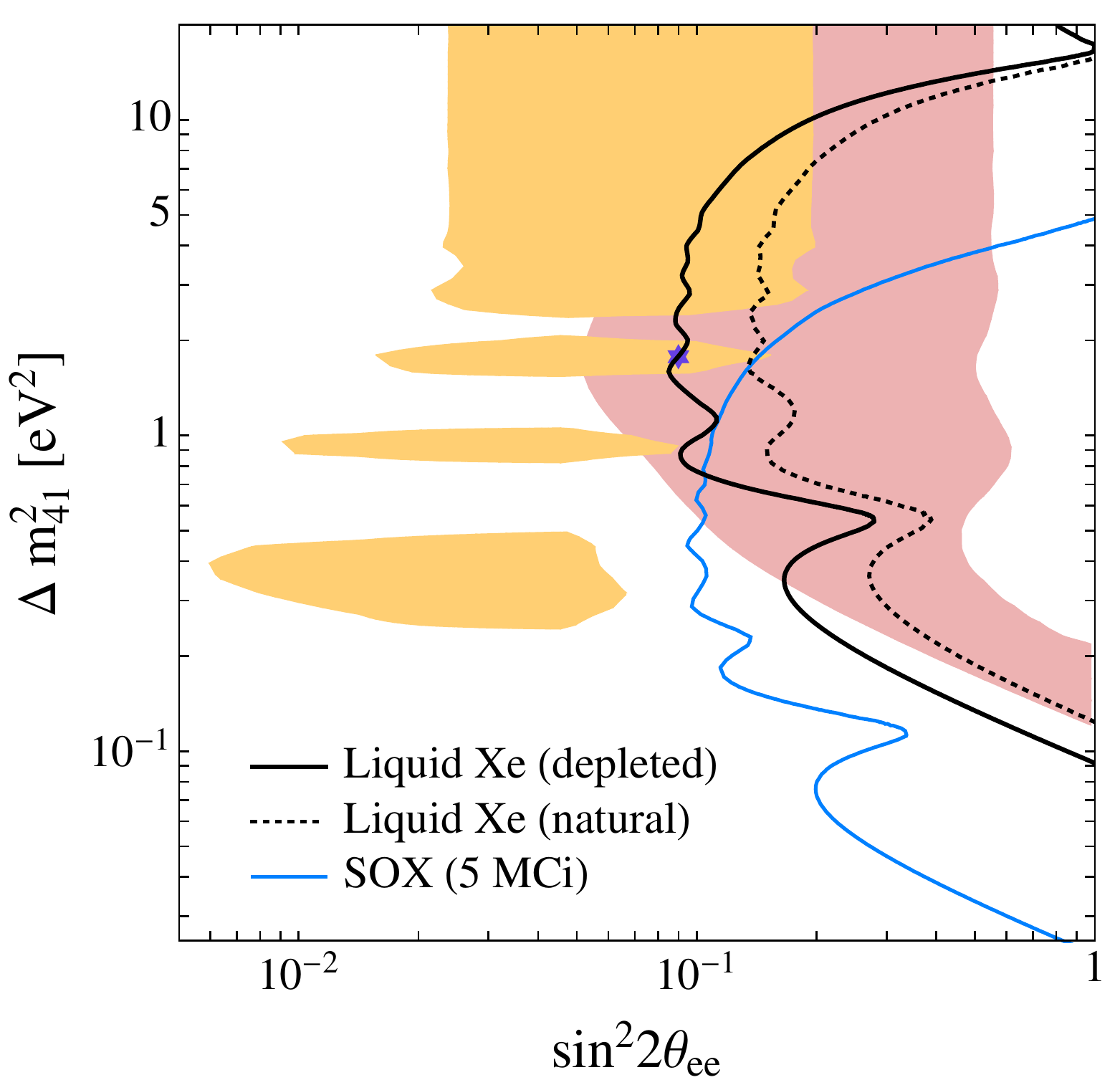} 
  \includegraphics[width=\columnwidth]{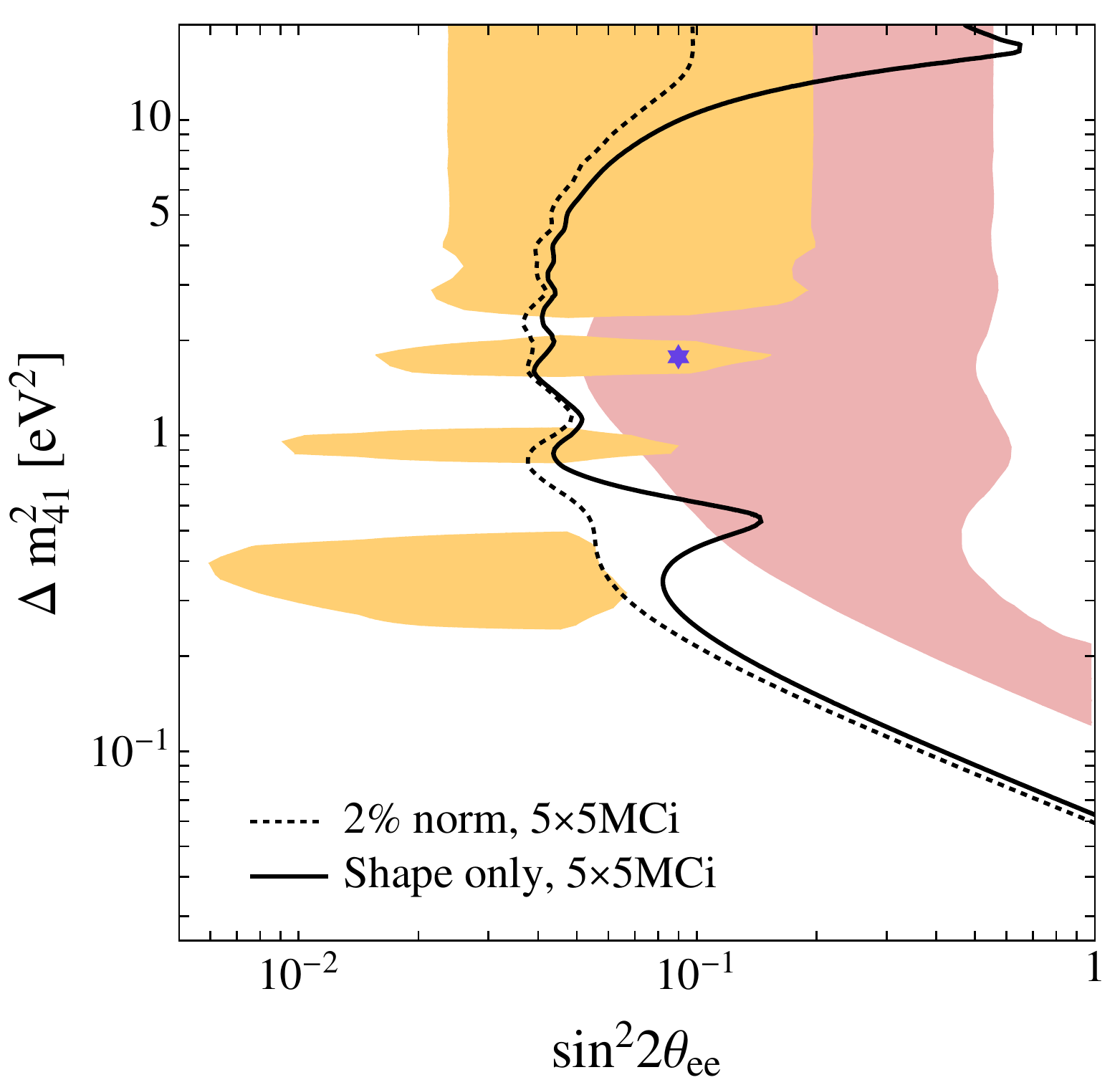}}
  \caption{Sensitivity to sterile neutrino oscillations as a function of 
    $\sin^22\theta_{ee}$ and $\Delta m^2_{41}$.  The parameter space to the 
    right of each line would be excluded at 95\% CL (2 d.o.f.). The shaded 
    areas show the 95\% CL allowed regions for the reactor (yellow) and
    gallium (pink) anomalies from a global fit to the 3+1 
    scenario, while the star indicates the best fit point from 
    a combined fit to both anomalies~\cite{Kopp:2013vaa}. Left panel: 
    expected sensitivity using shape information only (\textit{i.e.}, normalization is 
    left completely free). Black lines show the expected contours for the 
    LXe experiment described in the text. For comparison, the solid blue line 
    shows the SOX sensitivity using shape information alone. In both cases 
    a 5~MCi radioactive \Cr source is assumed. Right panel: expected 
    sensitivity for the LXe experiment with five source deployments. Results 
    in this panel are shown from analyses using shape information only (solid lines) and 
    shape plus normalization (dotted lines).}
\label{fig:sterile}
\end{figure*}

\section{Sterile neutrino searches}

A second possible application of the setup studied here would be the search 
for light sterile neutrinos. Sterile neutrinos arise in most models of neutrino 
mass generation. Searches for oscillations between active and sterile neutrinos 
have been conducted using many combinations of neutrino sources, detectors
and oscillation channels~\cite{Abazajian:2012ys}. Experimental
results are inconclusive, with strong tension between different data
sets (see, \textit{e.g.},
Refs.~\cite{Abazajian:2012ys,Giunti:2011hn,Kopp:2013vaa}).
In particular, an analysis of GALLEX and SAGE 
data shows an apparent deficit of events which is consistent with
oscillations involving a sterile neutrino with a squared mass 
difference with the active states $\Delta m^2 \sim 1\,\mathrm{eV}^2$ 
and a mixing angle such that $\sin^22\theta\sim0.1$~\cite{Acero:2007su}, 
which is commonly referred to as the \emph{gallium anomaly}. 
The smoking gun of a sterile neutrino
would be to observe events following an oscillating pattern in
$L/E_\nu$, the distance traveled by the neutrino divided by its
energy. A monochromatic neutrino source reduces the oscillating
pattern to a pure function of $L$.  Given the energy of the primary \Cr
neutrino the oscillation length at $\Delta m^2 = 1\, \mathrm{eV}^2$
would be $\sim\!$ 90~cm and thus the oscillating pattern would be
observable \emph{inside} a meter-scale detector.  The expected spatial
resolution in LXe TPCs is at the sub-cm level, which makes them ideal
candidates for such an experiment.

In our analysis, we have adopted a phenomenological approach based on a 3+1 framework, where there is only one extra sterile state at the eV scale. In this framework, the $\nu_e$ disappearance oscillation probability can be expressed in terms of one mixing angle and one mass squared splitting only, as:
\begin{equation}
P_{ee} = 1 - \sin^22\theta_{ee} \sin^2\left( \dfrac{\Delta m^2_{41} L}{4E_\nu}\right) \, ,
\end{equation} 
where $E_\nu$ is the neutrino energy, $L$ is the distance between the source and the interaction point in the detector, $\Delta m^2_{41} \equiv m^2_4 - m^2_1$ and $\theta_{ee}$ is an effective mixing angle. The sensitivity to sterile neutrino oscillations is computed using the
distance between the source and the interaction point ($L$) as the
main variable in the analysis. A binned Gaussian $\chi^2$ is built using 3~cm
wide bins. The detector is assumed to have a constant spatial
resolution of 1~cm; however, the largest uncertainty in $L$ comes from
the shape and size of the radioactive source itself.  In the present
work, we assume that the source will be a cylinder with both height and diameter 
equal to 14~cm.  To achieve such a compact source will require that the chromium 
material be enriched to $\sim$95\% in $^{50}$Cr. This is much higher than what was achieved by 
GALLEX~\cite{Cribier:1996cq}, but similar to what was reached by SAGE~\cite{Abdurashitov:1998ne}.
A smearing function was generated, by a Monte Carlo calculation, that simultaneously accounts
for the finite source size and the detector resolution. Two nuisance
parameters are added to the $\chi^2$ for signal and background
normalization uncertainties. A Gaussian penalty term (or pull-term) is
added to the $\chi^2$ for the background uncertainty and
marginalization is performed over both nuisance parameters. Unless
otherwise stated, no constraint is assumed for the flux normalization
and it is therefore left completely free during the fit. For the
backgrounds, an uncertainty of 0.5\% is assumed.  We expect this
could be achieved by using the data collected during the source-free
operation of the detector corresponding to the the dark matter search.
The process is repeated for each point in the $(\sin^22\theta, \Delta
m^2)$ parameter space.  Since the low-end threshold for the electron
recoil energy is not expected to have a great impact on the sterile
neutrino analysis, it is set to 5~keV in this case.

The results for the sterile neutrino sensitivity are shown in
Fig.~\ref{fig:sterile}. Since the sensitivity mainly comes from a
shape analysis, this measurement is far from being limited by
systematics. We find the main limiting factor to be the \Xe
background.  The expected number of background events (across the full
energy range) is around 51,130, from which around 44,000 are from
\Xe~\twobeta decays.  This background can be reduced by using LXe
depleted in $^{136}$Xe, though.  The sensitivities for both possibilities are
shown in Fig.~\ref{fig:sterile} (left).  For contrast, we also compare the
LXe sensitivity to that of Borexino/SOX~\cite{Borexino:2013xxa} using
a comparable, 5~MCi \Cr source (the SOX proposal is based on 
10~MCi). The complementarity in $\Delta m^2$ coverage of the two experiments is
evident.  In the case of LXe, the relatively high spatial resolution
improves sensitivity at high $\Delta m^2$, while for SOX, the
relatively large detector volume, or range in $L$, improves the reach
at low $\Delta m^2$.  It is remarkable that, for a depleted Xe
experiment, most of the region favored by the gallium anomaly is
covered with \emph{shape information alone}. Therefore, if the
gallium anomaly is correct this configuration would, with high
likelihood, confirm it by a clear observation of the
oscillatory pattern.

\section{Results for larger exposure}

Both the sterile neutrino and $\nu$MM searches are statistics limited
and would be improved by repeated redeployments of the \Cr source.
Repeated deployment has a precedent in GALLEX, which irradiated and
deployed the same source material twice~\cite{Hampel:1997fc}.  We will
consider five deployments, each identical in source strength and
duration to our previously considered single deployment.

In the case of the $\nu$MM search, the sensitivity with increased 
statistics is remarkable.  At 95\% CL, assuming a low-energy threshold 
of 2~keV, the corresponding bound is pushed down to $\mu_\nu = 3.31 
\times 10^{-12}\mu_B$.  If a lower threshold can be achieved, even the 
astrophysical limit would be surpassed.

The sensitivity to sterile neutrino oscillations with increased
statistics is shown in Fig.~\ref{fig:sterile} (right). In this case, we
assume 90\% depleted Xe, and show sensitivities from both a
shape-only analysis, and an analysis with the shape information
plus a 2\% uncertainty in the normalization.  As can be seen in the
comparison of the two bounds, no major improvement is expected from
imposing the normalization constraint, since the information comes
from observing the oscillating pattern in the detector.  Only in case
of large $\Delta m_{41}^2$, where the oscillation is averaged out,
would a normalization constraint help. According to our results, after
five deployments the full gallium anomaly as well as a sizable region
of the reactor anomaly would be covered at 95\% CL.

Obviously, with 10,000-50,000 $\nu_e e^-$ elastic scattering events
with momentum transfers in the 100\,keV range a number of precision
tests of the electro-weak sector of the SM become
possible. Assuming an absolute normalization at the 1\% level and
statistical errors at or below 1\%, overall accuracies for the total
cross section of 1-2\% appear feasible, providing constraints on the
weak mixing angle and its running.

\section{Summary and conclusions}

To summarize, our results indicate that the combination of a large
liquid xenon detector, designed and built to search for dark matter,
with a Mega-curie scale electron capture neutrino source would provide 
excellent reach in the search for the neutrino magnetic moment, exceeding 
the current laboratory bounds by at least a factor of 5.  With repeated 
source deployments, such an experiment would even be competitive with the 
best astrophysical limits.  Moreover, this combination would allow a test 
of the reactor and gallium anomalies; specifically, their interpretation 
as oscillations due to an eV-scale sterile neutrino. Its reach would be 
complementary to other source proposals with sensitivity to the oscillating 
pattern over almost the entire interesting range of $\Delta m^2$. Clearly, 
a detailed technical feasibility study is required, but so far no major 
technological obstacles have been identified.

\section*{Acknowledgments} 
This work has been supported by the U.S. Department of Energy under
award numbers \protect{DE-SC0003915} and \protect{DE-SC0009973}. We 
would like to thank the authors of Ref.~\cite{Kopp:2013vaa} for sharing 
with us their results for the global fit of the reactor and gallium 
anomalies.


%
\end{document}